\documentclass[fleqn, usenatbib]{mnras}
\usepackage{newtxtext, newtxmath} 	
\usepackage[T1]{fontenc}
\usepackage{graphicx}
\usepackage{amsmath}
\usepackage{mathtools}
\usepackage{multicol}
\usepackage{bm}
\usepackage{pdflscape}
\usepackage{natbib}
\usepackage[section]{placeins}
\usepackage{lipsum}
\usepackage{etoolbox}
\usepackage{tabularx}
\usepackage{xcolor}
\usepackage[toc, page]{appendix}
\usepackage{subfiles}
\usepackage{geometry}
\hypersetup{ 
	colorlinks		= true,
	urlcolor 		= blue,
	linkcolor 		= blue,
	citecolor 		= blue
}

\newcommand{\ddfrac}[2]{\frac{\displaystyle{#1}}{\displaystyle{#2}}}
\newcommand{\um}{\textsc{UniverseMachine}}
\newcommand{\mstar}{\ensuremath{\text{M}_\star}}
\newcommand{\msun}{\ensuremath{\text{M}_\odot}}
\newcommand{\refp}[1]{(\ref{#1})}

\providecommand{\noopsort}[1]{}

\title[Type Ia supernova rates]{Binaries drive high Type Ia supernova rates in
dwarf galaxies}

\author[J.W. Johnson, C.S. Kochanek \& K.Z. Stanek]{
	James W. Johnson,$^{1, 2}$\thanks{
		Contact e-mail: \href{mailto:johnson.7419@osu.edu}{johnson.7419@osu.edu}
	}
	Christopher S. Kochanek,$^{1, 2}$ and
	K. Z. Stanek$^{1, 2}$
	\\
	$^{1}$ Department of Astronomy, The Ohio State University,
	140 W. 18th Ave., Columbus, OH, 43210, USA
	\\
	$^{2}$ Center for Cosmology and Astroparticle Physics (CCAPP),
	The Ohio State University, 191 W. Woodruff Ave., Columbus, OH, 43210, USA
}

\date{Accepted XXX; Received YYY; in original form ZZZ}
\pubyear{2022}

\begin{document}
\label{firstpage}
\pagerange{\pageref{firstpage}--\pageref{lastpage}}
\maketitle

\begin{abstract}
The scaling of the specific Type Ia supernova (SN Ia) rate with host galaxy
stellar mass~$\dot{\text{N}}_\text{Ia} / \mstar \sim \mstar^{-0.3}$ as measured
in ASAS-SN and DES strongly suggests that the number of SNe Ia produced by a
stellar population depends inversely on its metallicity.
We estimate the strength of the required metallicity dependence by combining
the average star formation histories (SFHs) of galaxies as a function of their
stellar mass with the mass-metallicity relation (MZR) for galaxies and common
parametrizations for the SN Ia delay-time distribution.
The differences in SFHs can account for only~$\sim$30\% of the increase in the
specific SN Ia rate between stellar masses
of~$\mstar = 10^{10}$ and~$10^{7.2}~\msun$.
We find that an additional metallicity dependence of
approximately~$\sim$Z$^{-0.5}$ is required to explain the observed scaling.
This scaling matches the metallicity dependence of the close binary fraction
observed in APOGEE, suggesting that the enhanced SN Ia rate in low-mass
galaxies can be explained by a combination of their more extended SFHs and a
higher binary fraction due to their lower metallicities.
Due to the shape of the MZR, only galaxies below
$\mstar\approx3\times10^9~\msun$ are significantly affected by the
metallicity-dependent SN Ia rates.
The~$\dot{\text{N}}_\text{Ia} / \mstar \sim \mstar^{-0.3}$ scaling becomes
shallower with increasing redshift, dropping by factor of~$\sim$2
at~$10^{7.2}~\msun$ between~$z = 0$ and~$1$ with our~$\sim$$Z^{-0.5}$ scaling.
With metallicity-independent rates, this decrease is a factor of~$\sim$3.
We discuss the implications of metallicity-dependent SN Ia rates for one-zone
models of galactic chemical evolution.
\end{abstract}

\begin{keywords}
stars: supernovae -- stars: white dwarfs -- galaxies: abundances -- galaxies:
dwarf -- galaxies: evolution -- stars: mass function
\end{keywords}

\section{Introduction}
\label{sec:intro}

Type Ia supernovae (SNe Ia) arise from the thermonuclear detonation of a white
dwarf~\citep[WD;][]{Hoyle1960, Colgate1969}, the exposed carbon-oxygen core of
a low-mass star.
SN surveys have revealed that low-mass galaxies are more efficient producers
of these events than their higher mass counterparts~\citep[e.g.,][]{Mannucci2005,
Sullivan2006, Li2011, Smith2012}.
In particular,~\citet{Brown2019} found that the specific SN Ia rate -- the rate
per unit stellar mass -- scales approximately with the inverse square root of
the stellar mass itself ($\dot{\text{N}}_\text{Ia} / \mstar \sim \mstar^{-0.5}$)
using SNe Ia from the All-Sky Automated Survey for Supernovae
\citep[ASAS-SN;][]{Shappee2014, Kochanek2017} and assuming the~\citet{Bell2003}
stellar mass function (SMF).
The measurement depends on the SMF because the number of
observed SNe must be normalized by the number of galaxies in order to compute a
specific SN rate.
Consequently, the scaling becomes shallower ($\dot{\text{N}}_\text{Ia} / \mstar
\sim \mstar^{-0.3}$) when using the steeper~\citet{Baldry2012} double-Schechter
SMF parametrization~\citep{Gandhi2022}.
This change leads to agreement between the ASAS-SN measurements and Wiseman et
al.'s~\citeyearpar{Wiseman2021} estimates from the Dark Energy
Survey~\citep[DES;][]{DES2016} using the~\citet{Baldry2012} SMF.
Although the exact strength of the scaling depends on the SMF, it is clear
that the specific SN Ia rate is higher in dwarf galaxies.
There are a handful of potential pathways which could give rise to this
empirical result.
\par
First, the mean star formation histories (SFHs) of galaxies vary with the
stellar mass of the system.
In semi-analytic models of galaxy formation (see, e.g., the reviews of
\citealt{Baugh2006} and~\citealt{Somerville2015a}), dwarf galaxies in the field
have more extended SFHs than their higher mass counterparts.
This mass dependence is also seen in hydrodynamical simulations of galaxy
formation~\citep[e.g.,][]{GarrisonKimmel2019}.
Since SN Ia delay-time distributions (DTDs) decline with age, galaxies with
more recent star formation should have higher specific SN Ia rates.
\par
Second,~\citet{Kistler2013} argued that the dependence of the specific SN Ia
rate on stellar mass may be driven by metallicity.
Lower mass galaxies host lower metallicity stellar populations
\citep{Gallazzi2005, Kirby2013} and lower metallicity gas reservoirs
(\citealp{Tremonti2004};~\citealp*{Zahid2011};~\citealp{Andrews2013,
Zahid2014}).
\citet{Kistler2013} point out that lower metallicity stars leave behind higher
mass WDs which could potentially grow to the Chandrasekhar mass and
subsequently explode more easily than their less massive, high metallicity
counterparts.
Lower metallicity stars have weaker winds during the asymptotic giant branch
phase~\citep{Willson2000, Marigo2007}, leading to lower mass loss rates and
more massive cores~\citep{Kalirai2014}, producing more massive WDs for fixed
initial mass stars at lower metallcity (\citealp{Umeda1999};
\citealp*{Meng2008};~\citealp{Zhao2012}).
\par
Furthermore, in both the single~\citep[e.g.,][]{Whelan1973} and double
degenerate scenarios (e.g.,~\mbox{\citealp{Iben1984}};
\mbox{\citealp{Webbink1984}}), SNe Ia arise in binary systems.
Based on multiplicity measurements of Solar-type stars from the Apache Point
Observatory Galaxy Evolution Experiment~\citep[APOGEE;][]{Majewski2017},
\citet{Badenes2018} and~\citet*{Moe2019} find that the stellar close binary
fraction increases toward low metallicities.
Consequently, dwarf galaxies should have more potential SN Ia progenitors per
unit mass of star formation due to more massive WDs and a higher close binary
fraction.
Motivated by these results,~\citet{Gandhi2022} explore a handful of
parametrizations for the metallicity dependence of the SN Ia rate in
re-simulated galaxies from FIRE-2~\citep{Hopkins2018}.
They find that a~$Z^{-0.5}$ scaling where~$Z$ is the metallicity leads
to better agreement with the empirical relationship between galactic stellar
masses and stellar abundances than when using metallicity-independent SN Ia
rates.
\par
In this paper, we assume that the strong scaling of the specific SN Ia rate
with stellar mass is due to metallicity and conduct simple numerical
calculations to investigate its origin.
We combine the mean star formation histories of galaxies at fixed stellar
mass from the~\um~semi-analytic model~\citep{Behroozi2019} and the popular
$\tau^{-1}$ SN Ia DTD~\citep[e.g.,][]{Maoz2012a} with the mass-metallicity
relation (MZR) for galaxies~\citep{Tremonti2004, Andrews2013, Zahid2011,
Zahid2014}.
Given the mean SFH and DTD, we can compute the characteristic SN Ia rate for
galaxies of a given stellar mass, and by assuming that they lie along the
observed MZR, we can include various scalings of the rate with metallicity.
We describe the model in~\S~\ref{sec:galprops} and the effect on Type Ia rates
in~\S~\ref{sec:predictions}.
In~\S~\ref{sec:gce} we present simple models exploring the consequences of
metallicity-dependent SN Ia rates for galactic chemical evolution models.
We summarize our findings in~\S~\ref{sec:conclusions}.

\section{Galactic Properties}
\label{sec:galprops}

\begin{figure*}
\centering
\includegraphics[scale = 0.45]{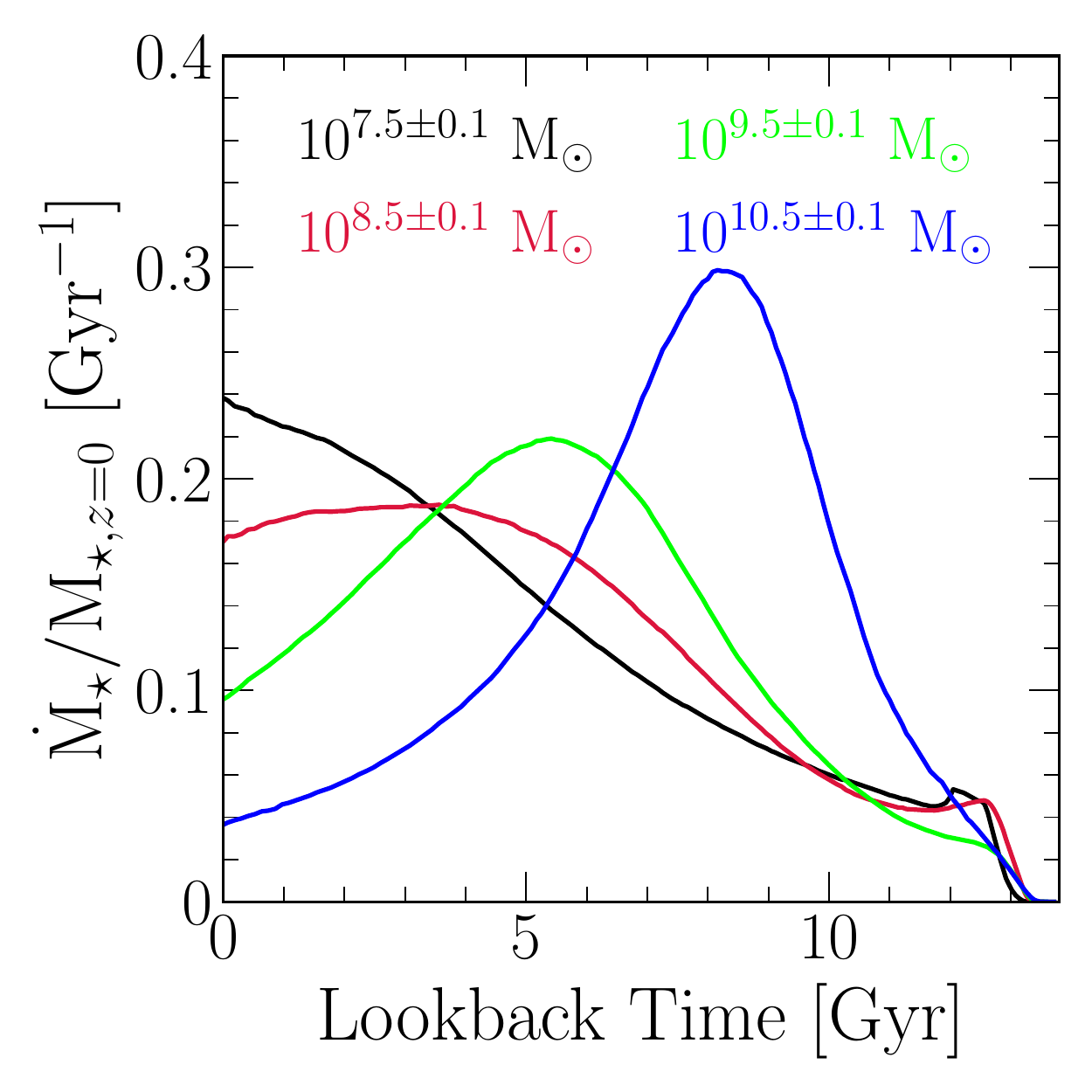}
\includegraphics[scale = 0.44]{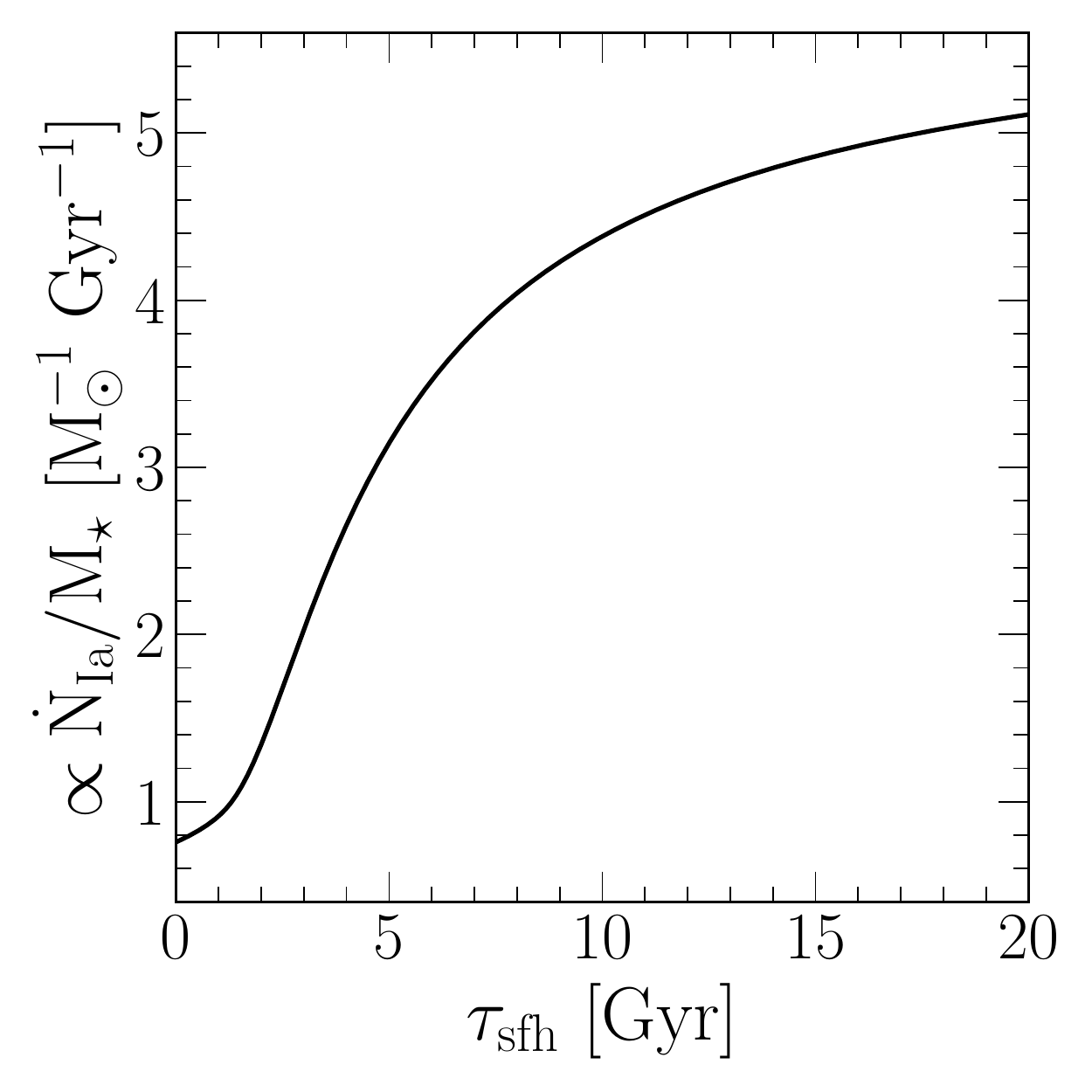}
\includegraphics[scale = 0.44]{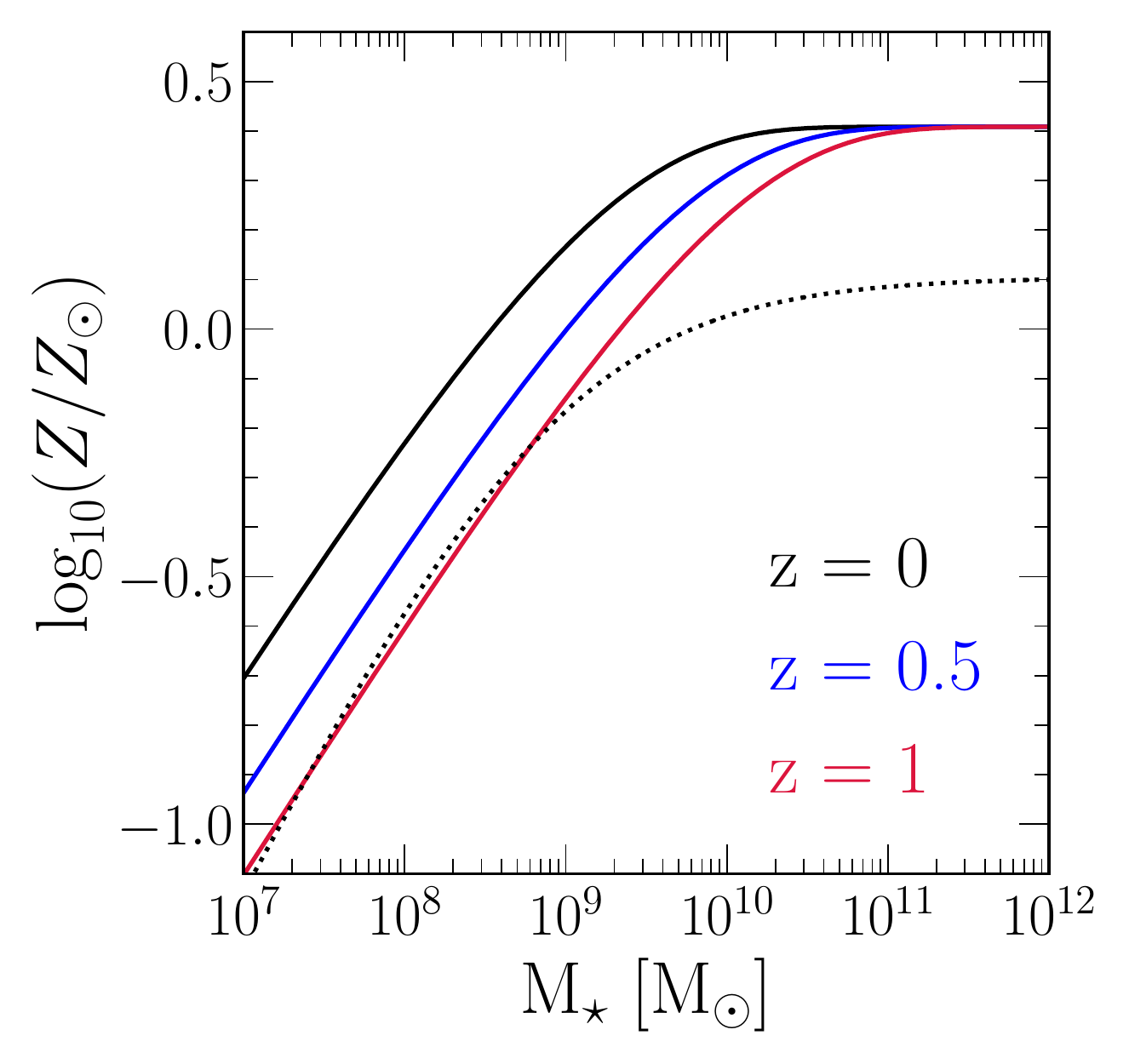}
\caption{
\textbf{Left}: The best-fit mean SFHs of the~\um~galaxies with present-day
stellar masses of $\mstar = 10^{7.5 \pm 0.1}$ (black),~$10^{8.5 \pm 0.1}$ (red),
$10^{9.5 \pm 0.1}$ (green), and $10^{10.5 \pm 0.1} \msun$ (blue) normalized by
their present-day stellar masses.
\textbf{Middle}: The specific SN Ia rate as a function of the e-folding
timescale of the SFH~$\tau_\text{sfh}$ assuming a linear-exponential time
dependence and a~$\tau^{-1}$ power-law SN Ia DTD.
\textbf{Right}: The redshift-dependent MZR reported by~\citet{Zahid2014} at
$z = 0$ (black solid),~$z = 0.5$ (blue), and~$z = 1$ (red).
For comparison, we include the~$z \approx 0$ MZR measured by
\citet[][black dotted]{Andrews2013}.
}
\label{fig:sfh_mzr}
\end{figure*}

We begin by examining how the mean galactic SFH varies with present-day stellar
mass as predicted by the~\um~semi-analytic model~\citep{Behroozi2019}.
Using dark matter halo properties supplied by the~\textit{Bolshoi-Planck} and
\textit{Multi-Dark Planck 2} dark matter only simulations~\citep{Klypin2016,
RodriguezPuebla2016},~\um~follows a conventional semi-analytic model framework
(see, e.g., the review in~\citealt{Somerville2015a}) and successfully
reproduces a broad range of well-constrained observables, including stellar
mass functions, cosmic SFRs, specific SFRs, quenched fractions, and UV
luminosity functions.
While some semi-analytic models have used the extended Press-Schechter
formalism~\citep{Press1974, Bond1991} to generate halo merger trees and push
the lower stellar mass limit of their model down to~$\mstar \approx 10^7~\msun$
\citep[e.g.][]{Somerville2015b}, an advantage of~\um~is that the high mass
resolution of the~\textit{Bolshoi-Planck} and~\textit{Multi-Dark Planck 2}
simulations allows merger trees down to~$\mstar = 10^{7.2}~\msun$ to be
obtained directly from the simulations.
Conveniently, this limit is approximately the lowest mass for which there are
empirical constraints on the specific SN Ia rate from ASAS-SN~\citep{Brown2019}
and DES~\citep{Wiseman2021}.
To relate these predictions to data from the untargeted ASAS-SN
survey~\citep{Shappee2014, Kochanek2017}, we take the full galaxy
sample from~\um, including both star forming and quenched galaxies as
well as both centrals and satellites, though centrals are the dominant
population across the full stellar mass range.
\par
In the left panel of Fig.~\ref{fig:sfh_mzr}, we show the best-fit mean SFH as a
function of lookback time in four narrow bins of present day stellar mass.
In general, low stellar mass galaxies have more extended SFHs than their
higher mass counterparts.
This effect is sufficiently strong that for stellar masses of
$\sim$$10^{7.5}~\msun$, typical SFRs are still increasing at the present day,
while~$\sim$$10^{10.5}~\msun$ galaxies experienced their fastest star formation
long ago.
\par
We adopt a DTD that scales with the age of a stellar population as~$\tau^{-1}$
starting at a delay time~$t_\text{D} = 100$ Myr as suggested by comparisons of
the cosmic SFH with the volumetric SN Ia rate as a function of redshift
(\citealp{Maoz2012a};~\citealp*{Maoz2012b};~\citealp{Graur2013, Graur2014}).
We conducted our analysis using alternative choices of the power-law
index as well as an exponential DTD with an e-folding timescale of
$\tau_\text{Ia} = 1.5$ Gyr and found similar conclusions in all cases.
We do not consider metallicity-dependent variations in the shape of the DTD
here, instead focusing on the overall normalization.
In principle, the minimum delay of the DTD could be as short as~$\sim$40 Myr if
WDs are produced by~$\lesssim$8~\msun~stars~\citep*[e.g.,][]{Hurley2000}, and
perhaps even shorter at low metallicity if the total metal content of a star
significantly impacts its lifetime~\citep[e.g.,][]{Kodama1997, Vincenzo2016}.
However, if SNe Ia require some additional time following WD formation, the
minimum delay will be longer.
Since we are interested in the first-order effects of variations in the SFH on
specific SN Ia rates, we assume a value of~$t_\text{D} = 100$ Myr.
In calculations using both~$t_\text{D} = 40$ Myr and~$t_\text{D} = 150$ Myr,
we found similar results.
\par
For an SFH~$\dot{M}_\star$ and DTD~$R_\text{Ia}$ as functions of lookback time
$\tau$, the specific SN Ia rate at a stellar mass~$\mstar$ is
\begin{equation}
\frac{\dot{N}_\text{Ia}(M_\star | \gamma)}{M_\star} \propto Z(M_\star)^\gamma
\ddfrac{
	\int_0^{T - t_\text{D}}\dot{M}_\star(\tau | M_\star) R_\text{Ia}(\tau) d\tau
}{
	\int_0^T \dot{M}_\star(\tau | M_\star) d\tau
}
\label{eq:specia}
\end{equation}
where~$T = 13.2$ Gyr is the time elapsed between the onset of star formation
and the present day.
To investigate the effects of metallicity, we add a power-law metallicity
scaling~$Z(\mstar)^\gamma$ where~$Z$ is given by the MZR.
We are only interested in the scaling of the rates with~\mstar, so we normalize
all rates to unity at~$\mstar = 10^{10}~\msun$ following~\citet{Brown2019}.
Although the denominator of equation~\refp{eq:specia} in detail should depend
on mass loss from stars as they eject their envelopes, this is an approximately
constant term which can safely be neglected in the interest of computing
relative rates ($\approx$40\% for a~\citealt{Kroupa2001} IMF; see discussion
in~\S\S~2.2 and 3.7 of~\citealt*{Weinberg2017}).
\par
To qualitatively illustrate how the specific SN Ia rate scales with the
timescale over which star formation occurs, we consider the simple example of a
linear-exponential
parametrization~$\dot{M}_\star \propto te^{-t/\tau_\text{sfh}}$ where
$t = T - \tau$.
The middle panel of Fig.~\ref{fig:sfh_mzr} shows equation~\refp{eq:specia} as
a function of the e-folding timescale~$\tau_\text{sfh}$ assuming~$\gamma = 0$.
The specific SN Ia rate is lowest in the limiting case of a single episode of
star formation (i.e.,~$\tau_\text{sfh} \rightarrow 0$), rises steeply until
$\tau_\text{sfh} \approx 10$ Gyr, and then flattens once
$\tau_\text{sfh} \gtrsim T$.
A higher specific SN Ia rate as observed in dwarf galaxies is therefore a
natural consequence of their more extended SFHs, though we demonstrate below
that this effect accounts for only a factor of~$\sim$2 increase in the rate
between~$10^{7.2}$ and~$10^{10} \msun$.
\par
The right panel of Fig.~\ref{fig:sfh_mzr} shows the MZR parametrized by
\citet[][see their equation 5]{Zahid2014}\footnote{
	We have transformed from their~$\log_{10}\text{(O/H)}$ measurements to the
	logarithmic abundance relative to the Sun~$\log_{10}(Z / Z_\odot)$ assuming
	the Solar oxygen abundance derived by~\citet{Asplund2009}.
} at redshifts~$z = 0$, 0.5 and 1 in comparison to the~\citet{Andrews2013}
parametrization at~$z = 0$.
Although~\um~allows us to investigate these effects at stellar masses as low as
$10^{7.2}~\msun$, the~\citet{Zahid2014} measurements are available only for
$\mstar \approx 10^9 - 10^{11}~\msun$ galaxies.
\citet{Andrews2013} used stacked spectra from the Sloan Digital Sky Survey
\citep[SDSS;][]{York2000} to obtain direct measurements of the oxygen abundance
in bins of stellar mass extending as low as~$\sim$$10^{7.4}~\msun$.
Relative to~\citet{Zahid2014}, the~\citet{Andrews2013} parametrization has a
lower plateau but otherwise a similar slope and turnover mass.
Because we simply normalize the rates to unity at~$\mstar = 10^{10}~\msun$,
only the shape of the MZR matters, and we find similar results using both
parametrizations.
In order to estimate SN Ia rates at redshifts of~$z = 0.5$ and~$z = 1$,
we use the redshift-dependent~\citet{Zahid2014} formalism
in~\S~\ref{sec:predictions}.
\par
Given a present-day stellar mass, we compute its SFH as a function of
lookback time by interpolating between the stellar mass and snapshot times
included in the~\um~predictions.
We then compute the specific SN Ia rate according to equation~\refp{eq:specia}
given the implied SFH and and a~$\tau^{-1}$ DTD, amplifying the rate by a
factor of~$Z^\gamma$ where the metallicity~$Z$ is computed from the
\citet{Zahid2014} MZR.
Because these calculations are simply using the~\um~SFHs, the results are
unaffected by the SMF dependence of the observational estimates (i.e.,
Eq.~\ref{eq:specia} can simply be divided by~$\mstar$ as opposed to an integral
over the SMF).

\section{Predicted SN Ia Rates}
\label{sec:predictions}

\begin{figure*}
\centering
\includegraphics[scale = 0.60]{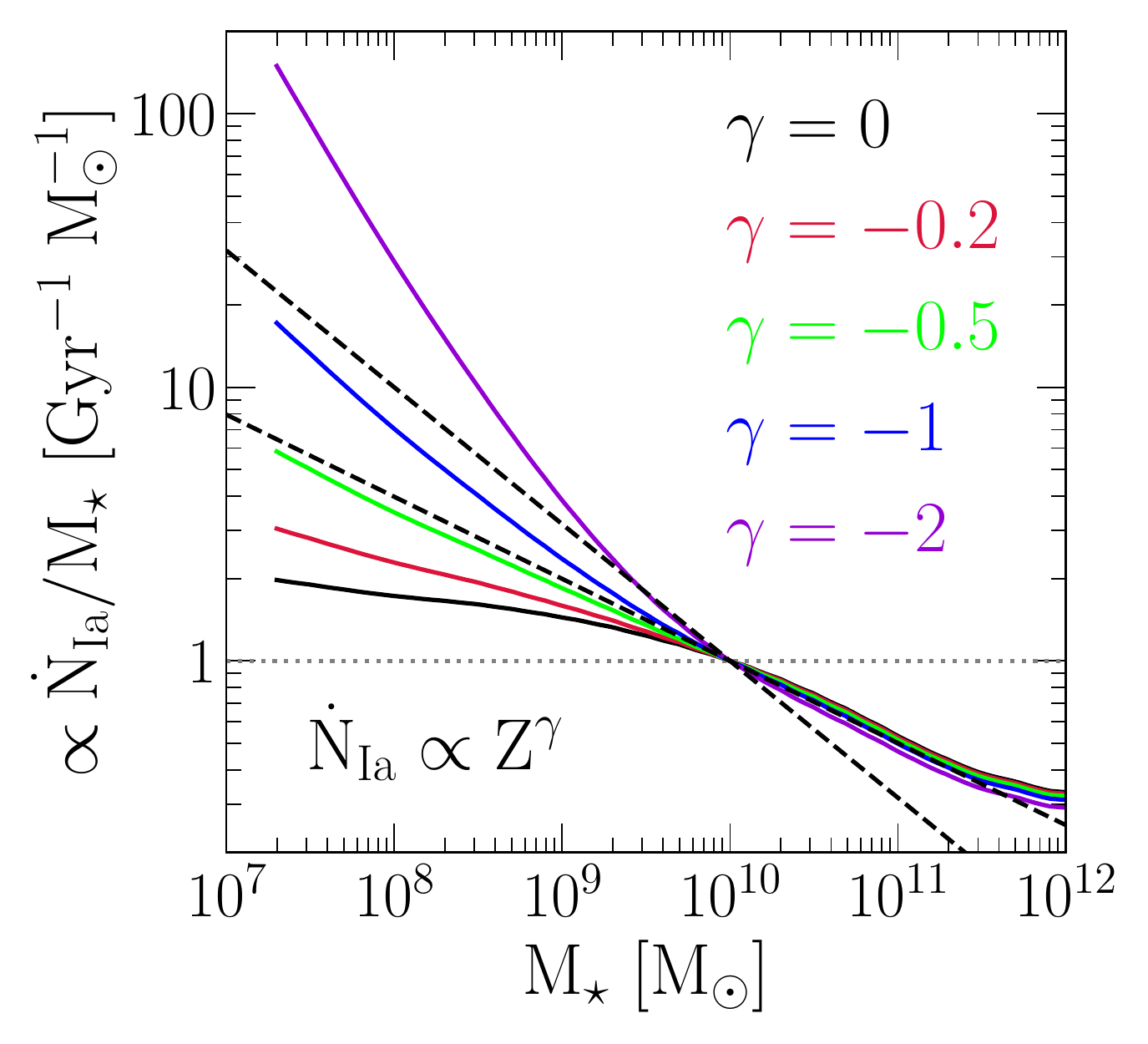}
\includegraphics[scale = 0.61]{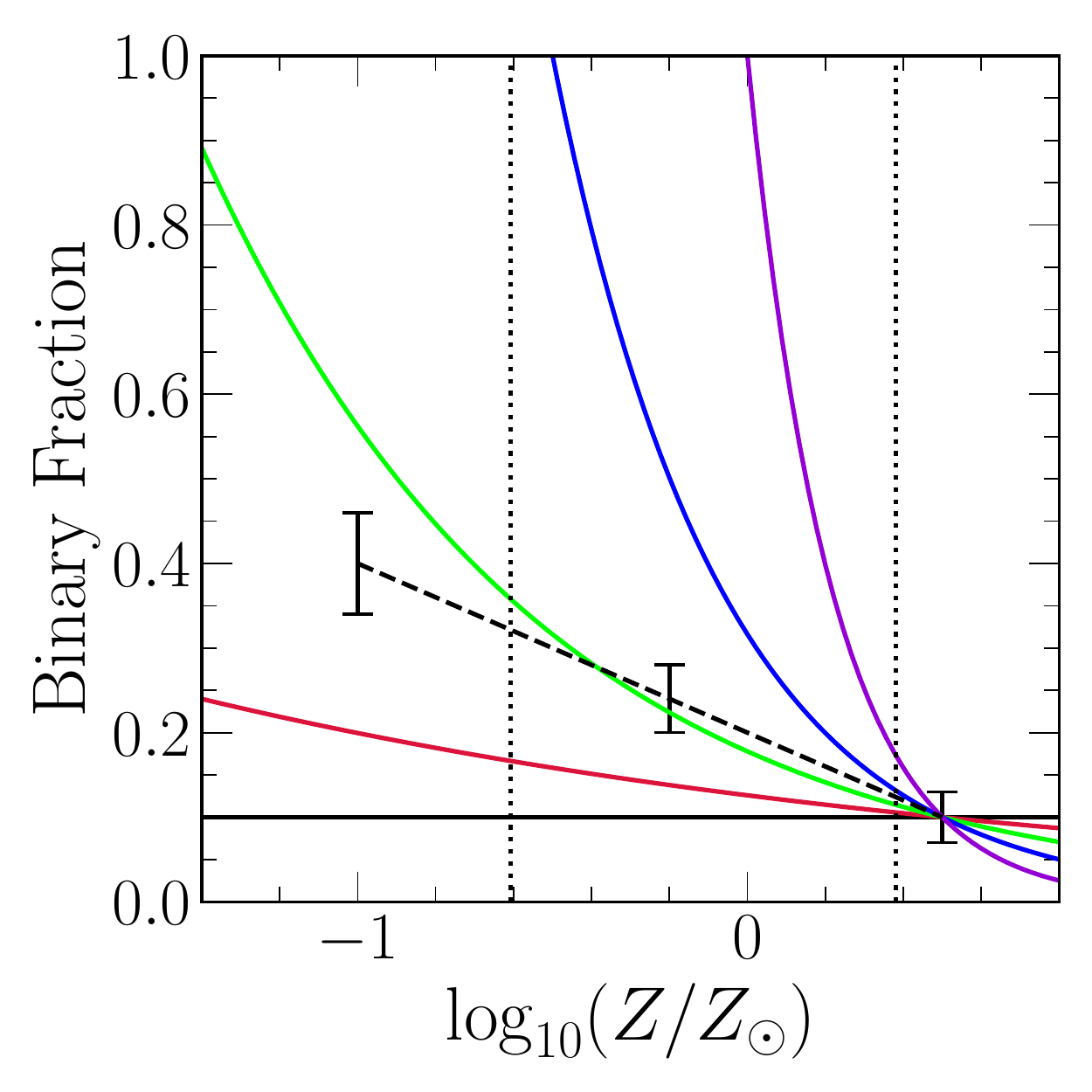}
\caption{
\textbf{Left}: Predicted scalings of the specific SN Ia rate with galaxy
stellar mass (see equation~\ref{eq:specia}) assuming the mean~\um~SFHs and a
single power-law~$Z^\gamma$ metallicity-dependence with~$\gamma = 0$ (i.e. no
dependence; black),~$\gamma = -0.2$ (red), $\gamma = -0.5$ (green),
$\gamma = -1$ (blue), and~$\gamma = -2$ (purple).
Following~\citet{Brown2019} and~\citet{Gandhi2022}, we normalize all rates to
a value of 1 at~$\mstar = 10^{10}~\msun$.
Black dashed lines denote the scalings of~$\dot{\text{N}}_\text{Ia} / \mstar
\sim \mstar^{-0.5}$ and~$\dot{\text{N}}_\text{Ia} / \mstar \sim \mstar^{-0.3}$
derived when normalizing the observed rates by the~\citet{Bell2003} and
\citet{Baldry2012} SMFs, respectively.
\textbf{Right}: The same metallicity scalings as in the left panel in
comparison to the close binary fractions observed in APOGEE
\citep[][black dashed line with error bars]{Moe2019} normalized to the observed
binary fraction of 10\% at~$\log_{10}(Z / Z_\odot) = +0.5$.
The characteristic metallicities of~$\mstar = 10^{7.2}~\msun$
($\log_{10}(Z / Z_\odot) \approx -0.6$) and~$10^{10}~\msun$ galaxies 
($\log_{10}(Z / Z_\odot) \approx +0.4$) are marked with black dotted lines.
}
\label{fig:specia_metdep}
\end{figure*}

\begin{figure*}
\centering
\includegraphics[scale = 0.65]{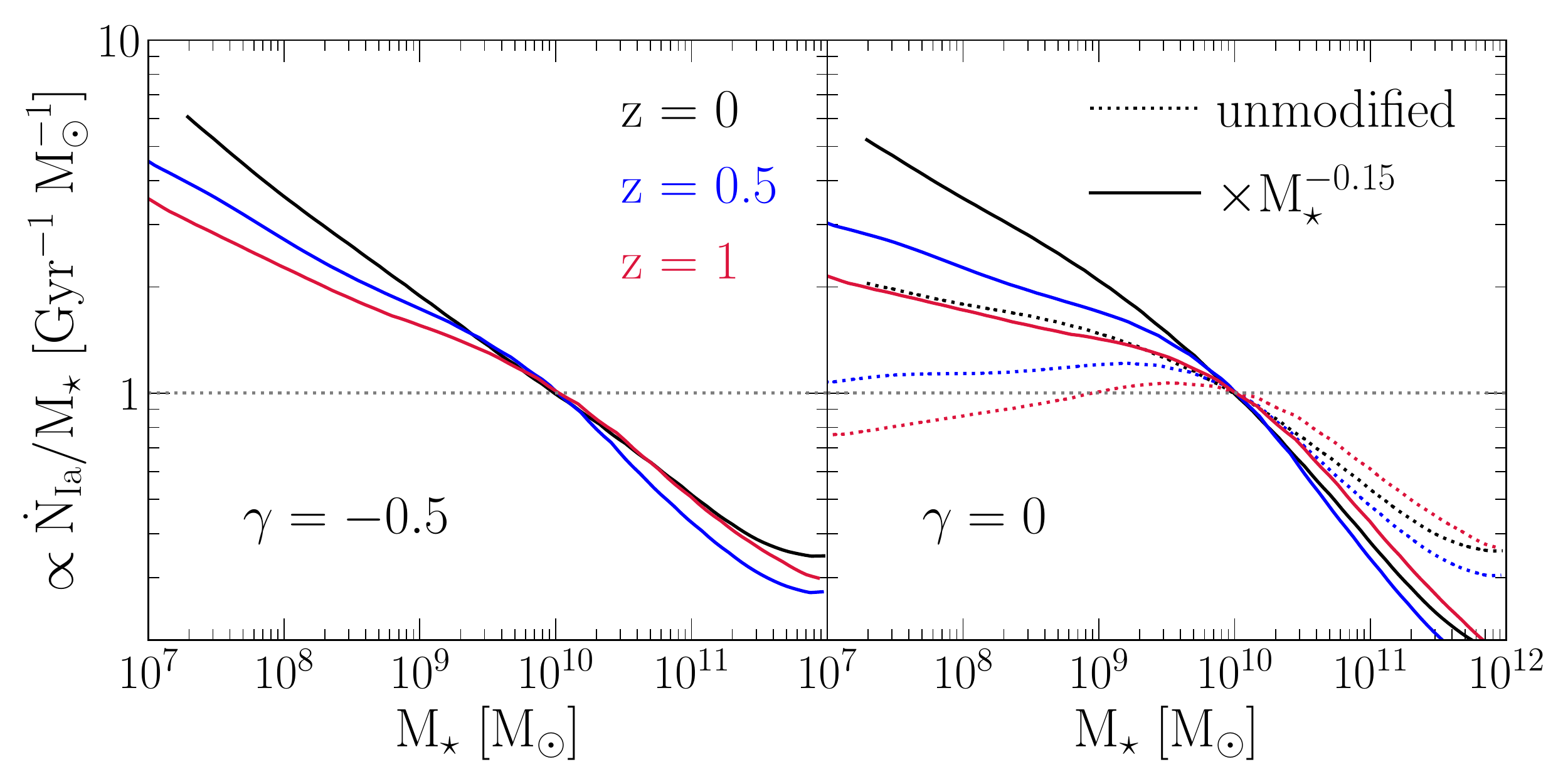}
\caption{
The specific SN Ia rate normalized to~$1$ at~$10^{10}~\msun$ as a function of
stellar mass with ($\gamma = -0.5$, left) and without a metallicity dependence
($\gamma = 0$, right) at redshifts $z = 0$ (black),~$z = 0.5$ (blue),
and~$z = 1$ (red).
In the right panel, we artificially scale the rates by a factor of
$(\mstar / 10^{10}~\msun)^{-0.15}$ to bring the~$z = 0$ predictions into
better agreement with an~$\sim \mstar^{-0.3}$ scaling as predicted by the
$\gamma = -0.5$ case.
Stellar masses correspond to the appropriate redshift (i.e., the rates for
$z = 1$ use the~$z = 1$ stellar masses and not the present-day stellar
masses).
We show the unmodified rates as dotted lines (the black solid line in the left
panel and the black dotted line in the right panel are the same as the green
line and black solid line in the left panel of Fig.~\ref{fig:specia_metdep}).
}
\label{fig:specia_zdep}
\end{figure*}

The left panel of Fig.~\ref{fig:specia_metdep} shows the specific SN Ia rate
as a function of~$\gamma$ in comparison to the~$\dot{\text{N}}_\text{Ia} /
\mstar \sim \mstar^{-0.5}$ and~$\dot{\text{N}}_\text{Ia} / \mstar \sim
\mstar^{-0.3}$ scalings of the observed rate with the~\citet{Bell2003} and
\citet{Baldry2012} SMFS, respectively.
The metallicity dependence has a significant impact only below
$\mstar \approx 3\times10^9~\msun$ due to the shape of the MZR;
this is the mass above which the MZR flattens considerably (see
Fig.~\ref{fig:sfh_mzr}).
Assuming no metallicity dependence (i.e.~$\gamma = 0$), these calculations
suggest that the variations in SFHs between~$\sim$$10^{7.2}$ and
$\sim$$10^{10}~\msun$ can account for only a factor of~$\sim$2 increase in the
specific SN Ia rate.
The~$\gamma = -0.5$ case is generally consistent with a mass dependence
of~$\mstar^{-0.3}$, while the steeper dependence of~$\mstar^{-0.5}$ would
require a stronger scaling of roughly~$\gamma \approx -1.5$.
\par
In the right panel of Fig.~\ref{fig:specia_metdep}, we compare the same
scalings to the close binary fractions in APOGEE measured by~\citet{Moe2019}.
The line at~$Z \approx 10^{-0.6} Z_\odot$ is the characteristic abundance of
an~$\mstar = 10^{7.2}~\msun$ galaxy in the~\citet{Zahid2014} parametrization.
For the range of metallicities spanned by the stellar masses we explore here,
the close binary fraction is remarkably consistent with a~$\gamma = -0.5$
scaling with metallicity.
If one instead takes~$Z \approx 0.1Z_\odot$ for a~$\sim$$10^{7.2}~\msun$ galaxy
as suggested by~\citet{Andrews2013}, then there is a slight tension between a
$\gamma = -0.5$ scaling and the close binary fraction measured by
\citet{Moe2019}.
There is some additional freedom to adjust the metallicity dependence beyond
that of binaries, so the agreement need not be perfect.
For example, any additional increase in the SN Ia rates not supplied by an
increased binary fraction could arise due to more massive WDs forming at
low~$Z$ -- the scenario postuled by~\citet{Kistler2013}.
Nonetheless, it appears that the scaling of the close binary fraction with
metallicity can explain the majority of the effect if rate scales with mass as
$\dot{\text{N}}_\text{Ia} / \mstar \sim \mstar^{-0.3}$.
If, instead, the~$\sim$$\mstar^{-0.5}$ scaling found using the~\citet{Bell2003}
SMF is accurate, then the required~$\gamma \approx -1.5$ scaling cannot be
explained by the close binary fraction alone as it would reach unphysical
values ($>1$) within the range of observed metallicities.
\par
Due to the evolution of the MZR, the mass dependence of the specific SN Ia rate
at different redshifts could empirically distinguish between~$\gamma = 0$ and
$\gamma = -0.5$.
To investigate this possibility, we simply evaluate equation~\refp{eq:specia}
over the appropriate range of lookback times assuming standard cosmological
parameters~\citep{Planck2014}.
For the~$\gamma = 0$ case, we also show the effect of applying an additional
$(\mstar / 10^{10}~\msun)^{-0.15}$.
This prefactor brings the~$\gamma = 0$ predictions into better agreement with
the empirical~$\dot{\text{N}}_\text{Ia} / \mstar \sim \mstar^{-0.3}$ scaling
at~$z = 0$ and is intended to encapsulate the mass scaling needed for some
other unknown process to sufficiently amplify the SN Ia rate at low stellar
masses if it is instead not due to metallicity effects.
\par
We show the resulting specific SN Ia rates as a function of stellar mass at
$z = 0$,~$0.5$ and~$1$ in Fig.~\ref{fig:specia_zdep}.
In both the~$\gamma = 0$ and $\gamma = -0.5$ cases, the scaling of the specific
SN Ia rate with galaxy stellar mass becomes shallower with increasing redshift.
If~$\gamma = -0.5$, these calculations suggest that it should decrease by a
factor of~$\sim$2 between~$z = 0$ and~$z = 1$ at~$\mstar \approx
10^{7.2}~\msun$, the lowest stellar mass for which we have made predictions at
all three redshifts.
If~$\gamma = 0$, then the rate instead decreases by a factor of~$\sim$3 at
$\sim 10^{7.2}~\msun$.
This difference arises because the metallicities of dwarf galaxies decrease
with increasing redshift and~$\gamma = -0.5$ allows them to sustain higher
SN Ia rates than if~$\gamma = 0$.
Empirically, the cosmic SN Ia rate increases with redshift
\citep[e.g.][]{Graur2014}, and we have verified that our framework reproduces
this result by integrating over the SMF (similar to equation
\ref{eq:hostmassdist} below).
Given this result and the lower stellar masses of the host galaxies at high
redshift, one might expect the trend to steepen with increasing~$z$.
The slope instead decreases here because the lines in Fig.~\ref{fig:specia_zdep}
(right panel) are moving to the right with time as galaxies grow in mass, and
we normalize to unity at a stellar mass of~$10^{10}~\msun$ at all redshifts.

\begin{figure}
\centering
\includegraphics[scale = 0.55]{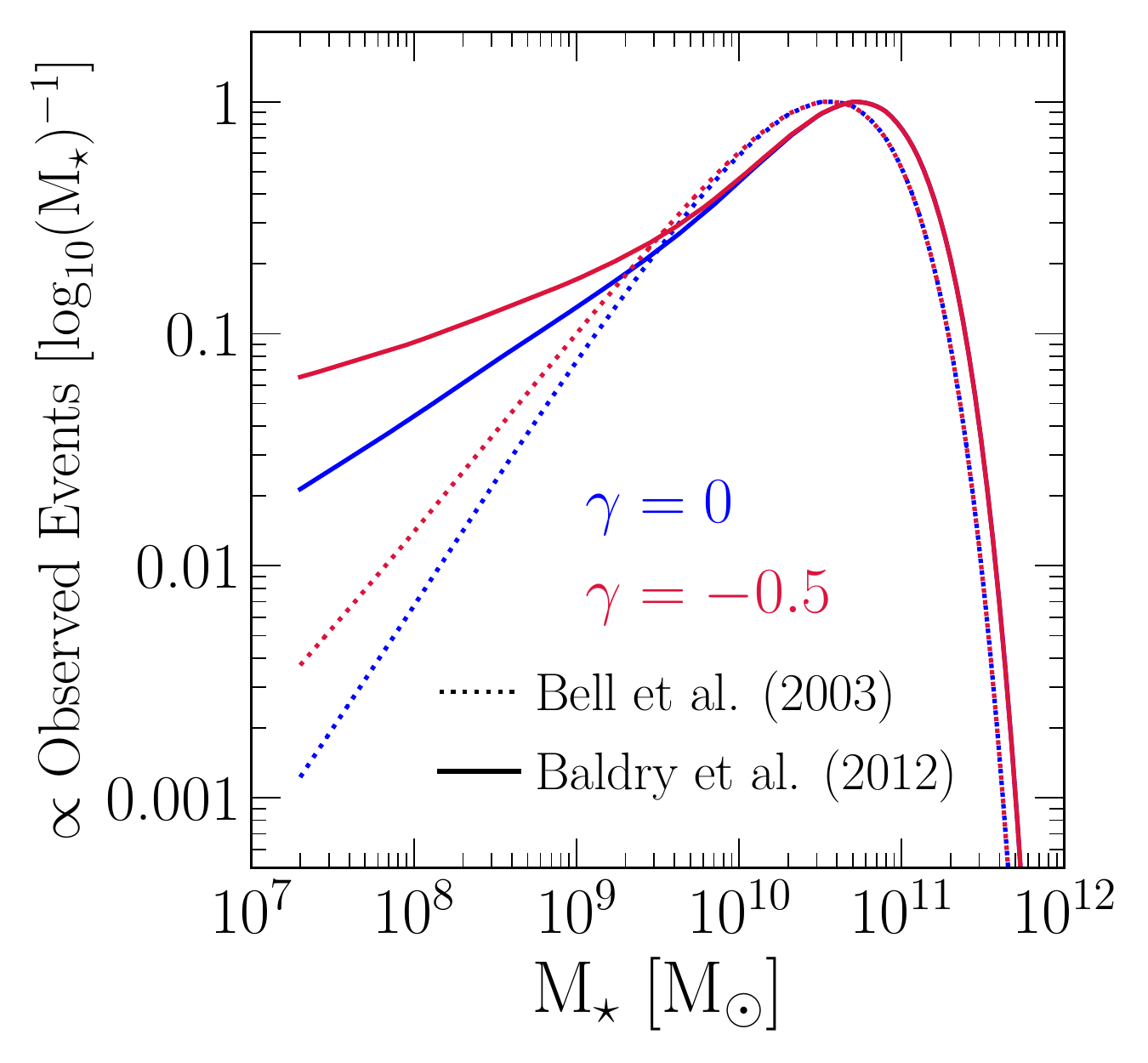}
\caption{
The predicted stellar mass distribution of~$z = 0$ SN Ia host galaxies in an
untargeted survey (see equation~\ref{eq:hostmassdist}) with the
\citet[][dotted]{Bell2003} and~\citet[][solid]{Baldry2012} SMFs for both
metallicity-dependent ($\gamma = -0.5$, red) and metallicity-independent rates
($\gamma = 0$, blue).
All distributions are normalized to a maximum value of 1.
}
\label{fig:hostmassdist}
\end{figure}

While empirical measurements of the specific SN Ia rate as a function of
stellar mass depend on the assumed SMF~\citep{Gandhi2022}, the host galaxy
mass distribution of observed events does not, making it a potentially more
observationally feasible diagnostic.
As noted in Fig.~\ref{fig:specia_metdep}, only dwarf galaxies are
significantly affected by a metallicity-dependent scaling of SN Ia rates due to
the shape of the MZR, so a~$\gamma \approx -0.5$ scaling should appear as an
enhanced SN Ia rate at the low-mass end of the distribution.
Although this empirical measurement does not depend on the SMF, our theoretical
prediction does because we must take into account the relative abundances of
galaxies of different stellar masses.
The observed rate in a bin of stellar mass can be expressed as the product
of the characteristic rate~$\dot{\text{N}}_\text{Ia}$ at a given stellar mass
and the integral of the SMF~$\Phi(\mstar)$ over the bin in stellar mass,
\begin{equation}
\dot{N}_\text{Ia,cosmic}(M_\star | \gamma) = \dot{N}_\text{Ia}(M_\star | \gamma)
\int_{M_\star}^{M_\star + dM_\star} \Phi(M_\star) dM_\star,
\label{eq:hostmassdist}
\end{equation}
where~$\dot{\text{N}}_\text{Ia}$ is the numerator of equation~\refp{eq:specia}.
We show this distribution in Fig.~\ref{fig:hostmassdist} for each combination
of~$\gamma = 0$ and~$-0.5$ and the~\citet{Bell2003} and~\citet{Baldry2012} SMFs,
normalizing to a maximum value of unity.
For untargeted surveys like ASAS-SN, equation~\refp{eq:hostmassdist} should
describe the observed host galaxy stellar mass distribution exactly, whereas
targeted surveys like the Lick Observatory SN Search~\citep[LOSS;][]{Li2000,
Filippenko2001} would need to correct for their target galaxy selection
criteria.
\par
Fig.~\ref{fig:hostmassdist} shows that galaxies with stellar masses of
$\mstar = 10^{10} - 10^{11}~\msun$ should dominate the SN Ia rate for all
choices of the SMF and~$\gamma$.
This peak rate simply represents the galaxies that dominate the stellar mass.
For a given choice of the SMF, $\gamma = -0.5$ increases the number of SNe Ia
at~$\mstar \approx 10^{7.2}~\msun$ by a factor of~$\sim$3 relative to
$\gamma = 0$.
However, Fig.~\ref{fig:hostmassdist} also shows that the enhancement is small
compared to the differences between the two SMFs.
Between the peak and~$10^{7.2}~\msun$, the ~\citet{Bell2003} host mass
distribution drops by~$\sim$2.5 orders of magnitude while
that of~\citet{Baldry2012} drops by~$\sim$1.5 orders of magnitude.
Fig.~\ref{fig:hostmassdist} illustrates the need for precise knowledge of the
SMF to accurately determine the metallicity-dependence of SN Ia rates.

\section{Galactic Chemical Evolution}
\label{sec:gce}

The realization that SN Ia rates likely depend on metallicity with a
$\gamma = -0.5$ dependence has important implications for galactic chemical
evolution models, which typically assume metallicity-independent rates.
To demonstrate this, we briefly explore several one-zone models based
on~\citet{Johnson2020} which predict the evolution of O (produced only in
massive stars) and Fe (produced in both massive stars and SNe Ia).
We use an exponential SFH with an e-folding timescale of~$\tau_\text{sfh} = 6$
Gyr and a minimum delay of~$t_\text{D} = 100$ Myr before the onset of SNe Ia
from a given stellar population.
For metallicity-dependent rates, we simply apply a~$(Z / Z_\odot)^{-0.5}$
prefactor to their Fe yield of~$y_\text{Fe}^\text{Ia} = 0.0017$, which assumes
that the shape of the DTD does not vary with metallicity -- only the
normalization.
Otherwise, these models are the same as~\citet{Johnson2020}.
\par
\begin{figure*}
\centering
\includegraphics[scale = 0.46]{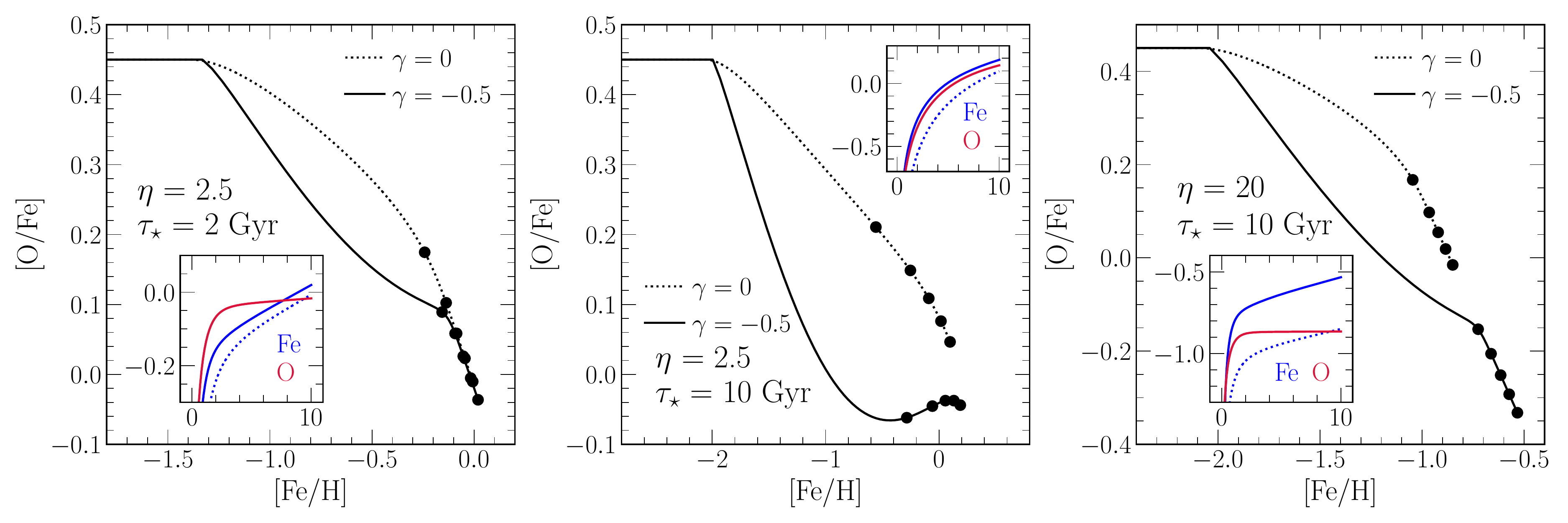}
\caption{
A comparison of one-zone galactic chemical evolution models based on
\citet[][for details, see their~\S~2]{Johnson2020} with ($\gamma = -0.5$,
solid) and without ($\gamma = 0$, dotted) metallicity-dependent SN Ia rates.
Tracks denote the O and Fe abundances in the interstellar medium parametrized
as a function of time with points marked at~$T = 2$, 4, 6, 8, and 10 Gyr.
Insets illustrate [O/H] and [Fe/H] as a function of time in Gyr for the
corresponding model.
We note on each panel the choice of the outflow mass-loading factor~$\eta$ and
the star formation efficiency timescale~$\tau_\star$.
}
\label{fig:onezone_app}
\end{figure*}

Fig.~\ref{fig:onezone_app} illustrates the predictions of this SFH with
$(\eta, \tau_\star) = (2.5, 2~\text{Gyr}), (2.5, 10~\text{Gyr})$, and
$(20, 10~\text{Gyr})$ where~$\eta \equiv \dot{M}_\text{out} / \dot{M}_\star$ is
the mass-loading factor describing the efficiency of outflows and
$\tau_\star \equiv M_\text{gas} / \dot{M}_\star$ is the inverse of the
star formation efficiency.
These values are appropriate for the Solar neighbourhood with efficient (left
panel) and inefficient (middle panel) star formation and for a dwarf galaxy
(right panel).
In all cases,~$\gamma = -0.5$ predicts a much more abrupt descent from the
high [O/Fe] plateau because of the higher Fe yield at low metallicity.
If incorporated into GCE models, this could impact the inferred evolutionary
timescales of the thick disc, known to host many of the high [O/Fe] stars in
the Milky Way~\citep{Hayden2017}.
When the equilibrium abundance is near-Solar but star formation is slow (middle
panel), the model predicts a ``secondary plateau'' in [O/Fe] because the Fe
enrichment rate slows down due to the metallicity-dependence of the yield
(see inset).
This is a noteworthy theoretical prediction because generally
[$\alpha$/Fe] and [Fe/H] reach equilibrium at similar times and no secondary
plateau arises~\citep[e.g.,][]{Weinberg2017}.
\par
These models are intended to be illustrative rather than quantitative.
In general, the only regions of chemical space where~$\gamma = 0$ and
$\gamma = -0.5$ agree are at near-Solar abundances and along the high [O/Fe]
plateau, which occurs before the onset of SN Ia enrichment.
A~$\gamma = -0.5$ scaling has the strongest impact at low~$Z$, where the
metallicity-dependence of the yield shifts the Fe abundances by~$\sim$0.5 dex
in this example.

\section{Discussion and Conclusions}
\label{sec:conclusions}

Building on LOSS~\citep{Li2011},~\citet{Brown2019} and~\citet{Wiseman2021}
found the SN Ia rates rise steeply toward low stellar mass.
The exact slope depends on the adopted SMF, with
$\dot{\text{N}} / \mstar \sim \mstar^{-0.3}$ for~\citet{Baldry2012} and
$\dot{\text{N}} / \mstar \sim \mstar^{-0.5}$ for~\citet{Bell2003}.
To explain this scaling with mass, we use the mean SFHs of galaxies predicted
by the~\um~\citep{Behroozi2019} semi-analytic model of galaxy formation, a
standard~$\tau^{-1}$ DTD~\citep[e.g.,][]{Maoz2012a}, and the empirical MZR as
parametrized by~\citet{Zahid2014} to relate stellar mass to metallicity and
build-in a~$Z^\gamma$ SN Ia rate dependence.
Our results depend only on the shape of the MZR and not its absolute
calibration.
While lower mass galaxies have younger stellar populations, we find that this
accounts for only a factor of~$\sim$2 increase in the specific rate between
$\mstar = 10^{7.2}$ and~$10^{10}~\msun$.
We can match the~$\mstar^{-0.3}$ increase if~$\gamma \approx -0.5$,
but~$\gamma \approx -1.5$ is required to explain the steeper
$\dot{\text{N}} / \mstar \sim \mstar^{-0.5}$ scaling.
\par
A scaling of~$\gamma = -0.5$ is in excellent agreement with the dependence of
the close binary fraction measured in APOGEE, which increases from~$\sim$10\%
at~$\sim$$3Z_\odot$ to~$\sim$40\% at~$\sim$$0.1Z_\odot$~\citep{Moe2019}.
This close match suggests that if a scaling of
$\dot{\text{N}}_\text{Ia} / \mstar \sim \mstar^{-0.3}$ is accurate, then the
elevated SN Ia rates in dwarf galaxies can be explained by a combination of
their more extended SFHs and an increased binary fraction compared to their
higher mass counterparts due to differences in metallicity.
While~\citet{Gandhi2022} motivate their investigation from this viewpoint, here
we take this argument one step further and postulate that this accounts for
the~\textit{entire} increase in the specific SN Ia rate because the binary
fraction can naturally account for a factor of~$\sim$3 increase over
the~$\sim$1 decade in metallicity spanned by~$\mstar = 10^{7.2} - 10^{10}
\msun$ galaxies.
The suggestion by~\citet{Kistler2013} that the increased WD mass at low
metallicities drives the effect is likely a subdominant effect.
Higher mass WDs for lower metallicity may be an important component if the
steeper~$\sim \mstar^{-0.5}$ scaling inferred with the~\citet{Bell2003} SMF is
correct.
\par
At first glance, an inverse dependence of SN Ia rates on metallicity may seem
at odds with the results in~\citet{Holoien2022} finding that dwarf galaxy
hosts of ASAS-SN SNe Ia tend to be oxygen-rich relative to similar mass
galaxies.
However, SN hosts are likely not a representative sample of the underlying
galaxy population because the steeply declining DTD~\citep[e.g.,][]{Maoz2012a}
means that the intrinsically highest SN Ia rates at any mass should be in
systems which experienced a recent starburst ($\lesssim$1 Gyr ago).
Since oxygen is produced by massive stars with short lifetimes
\citep*[e.g.,][]{Hurley2000, Johnson2019}, these galaxies should also have a
higher-than-average oxygen abundance~\citep[see, e.g.,][]{Johnson2020}.
In other words, SN Ia hosts at fixed mass should be more metal-rich than the
average galaxy.
\par
The calculations we have presented here are simplified in several regards.
We assumed the characteristic SFH predicted by a semi-analytic model of
galaxy formation at all stellar masses.
Our parametrization of the MZR includes no intrinsic scatter, and taking the
\citet{Zahid2014} MZR at face value for use in a power-law scaling implicitly
assumes that all SNe Ia arise from stellar populations near the gas-phase
abundance.
Although in principle galaxies populate distributions of finite width in each
of these quantities, these approximations should be fine for the purposes of
predicting average trends.
\par
Although current surveys lack the depth required to pin down SN rates across
multiple decades of stellar mass at~$z = 1$, the sample sizes necessary to do
so may be available from next-generation facilities.
First and foremost, the Nancy Grace Roman Space Telescope~\citep{Spergel2013,
Spergel2015} will obtain large samples of SNe.
Roman has excellent prospects for discovering all classes of SNe at redshifts
as high as~$z \gtrsim 2$ and beyond~\citep{Petrushevska2016}.
The difficulty in empirically constraining the specific SN Ia rate at
$z \approx 1$ instead comes from uncertainties in the SMF.
Even at~$z = 0$, these measurements are difficult due to the flux-limited
nature of most surveys and the broad range of luminosities and mass-to-light
ratios spanned by galaxies~\citep[see the discussion in][]{Weigel2016}.
Between~$10^{7.2}$ and~$10^{10}~\msun$, the factors of 2 and 3 predicted by our
calculations with~$\gamma = -0.5$ and $\gamma = 0$ are produced by power-law
indices of~$-0.108$ and~$-0.170$, respectively.
The difference between the two (0.062) is the minimum precision required for
the scaling of the SMF at the low-mass end -- only slightly larger than the
precision achieved by~\citet[][$\pm 0.05$, see their Fig. 13]{Baldry2012}.
This empirical test therefore requires at least their level of precision but
at~$z \approx 1$.
\par
A metallicity dependence of~$Z^{-0.5}$ strongly impacts the evolution of Fe
in one-zone models of galactic chemical evolution.
The considerable impact that a~$\gamma = -0.5$ scaling has on the predictions
indicates that evolutionary parameters inferred from one-zone model fits to
multi-element abundance ratios may need revised.
The strongest impact is for dwarf galaxies, where the abundances are low and
the higher yields predict substantial shifts in the position of the
evolutionary track in abundance space.

\section{Acknowledgments}
\label{sec:acknowledgments}

We are grateful to David H. Weinberg for valuable comments which improved this
manuscript.
JWJ thanks Todd A. Thompson, Jennifer A. Johnson, Adam K. Leroy, and the rest
of the Ohio State University Gas, Galaxies, and Feedback group for valuable
discussion.
JWJ also acknowledges financial support from an Ohio State University
Presidential Fellowship.
CSK and KZS are supported by NSF grants AST-1814440 and AST-1908570.
ASAS-SN is funded in part by the Gordon and Betty Moore Foundation through
grants GBMF5490 and GBMF10501 to the Ohio State University, and also funded in
part by the Alfred P. Sloan Foundation grant G-2021-14192.
\par\null\par\noindent
\textit{Software}:~\textsc{NumPy}~\citep{Harris2020},~\textsc{SciPy}
\citep{Virtanen2020},~\textsc{Matplotlib}~\citep{Hunter2007},
\textsc{Astropy}~\citep{Astropy2013, Astropy2018, Astropy2022}.

\section{Data Availability}
The only data in this paper are the mean star formation histories of
the~\um~galaxies, publicly available
at~\url{https://www.peterbehroozi.com/data.html}.

\bibliographystyle{mnras}
\bibliography{ms}

\label{lastpage}
\end{document}